# Domain wall motion in epitaxial Pb(Zr,Ti)O$_3$ capacitors investigated by modified piezoresponse force microscopy

S. M. Yang, J. Y. Jo, D. J. Kim, H. Sung, and T. W. Noh[*]

*ReCOE & FPRD, Department of Physics and Astronomy, Seoul National University, Seoul 151-747, Korea*

H. N. Lee[**]

*Materials Science and Technology Division, Oak Ridge National Laboratory, Oak Ridge, Tennessee 37831, USA*

J.-G. Yoon

*Department of Physics, University of Suwon, Suwon, Gyunggi-do 445-743, Korea*

T. K. Song

*School of Nano and Advanced Materials Engineering, Changwon National University, Changwon, Gyeongnam 641-773, Korea*

We investigated the time-dependent domain wall motion of epitaxial PbZr$_{0.2}$Ti$_{0.8}$O$_3$ capacitors 100 nm-thick using modified piezoresponse force microscopy (PFM). We obtained successive domain evolution images reliably by combining the PFM with switching current measurements. We observed that domain wall speed ($v$) decreases with increases in domain size. We also observed that the average value of $v$, obtained under applied electric field ($E_\text{app}$), showed creep behavior: *i.e.* $<v> \sim \exp(-E_0/E_\text{app})^\mu$ with an exponent $\mu$ of 0.9 ± 0.1 and an activation field $E_0$ of about 700 kV/cm.

It is important, both scientifically and technologically, to understand domain wall motion during polarization switching in ferroelectric (FE) thin films. There have been extensive studies on FE polarization switching dynamics by measuring polarization-electric field hysteresis,[1,2] switching currents,[3] and microscopic images of FE domains.[4-13] However, most measurements have serious limitations in providing the information required to understand time-dependent domain wall motion during the switching process. The FE domain wall width is typically 1 nm or less,[14] and the domain size typically varies from a few nm to above 1 $\mu$m. In addition, FE polarization switching is known to occur on a timescale between nanoseconds and seconds. Therefore, the study on domain wall dynamics in FE thin films poses a challenge due to the lengths and timescales involved.

Piezoresponse force microscopy (PFM) is commonly used to investigate images of FE domains in the nanoscale region, especially under static conditions. Tybell *et al.* extended the PFM technique to observe domain wall motion under various applied electric fields ($E_{app}$).[4] Using a conductive PFM tip, they applied $E_{app}$ to a film. By varying time ($t$) duration of the electric pulse, they created a nucleus with opposite polarization and studied how the reversed domain grows as a function of $t$. Similar PFM methods have been used extensively by other workers.[5-9]

Although this conventional PFM approach is rather simple and convenient to use, it has several important drawbacks in investigating domain wall motion inside real FE devices. First, $E_{app}$ is strongly dependent on the tip conditions, such as tip shape, tip radius, and the relative distance between the tip and film, some of which are quite difficult to control. Second, $E_{app}$ is not uniform, and so a rather complicated analysis is required to estimate the accurate $E_{app}$ values. Third, it can observe only local switching starting from the nucleus just below the tip, generated by a strong $E_{app}$. However, in most FE devices, that have capacitor geometry, domain switching begins to occur from nuclei created inhomogeneously at numerous sites near the interfaces between the film and the electrodes under rather uniform external field.

The limitations of the conventional PFM approach can be overcome by using top electrode and measuring piezoresponse signals coming from the FE capacitor.[15] Using thin film FE capacitors, Gruverman *et al*. employed the step-by-step switching approach in conjunction with PFM images.[10] They directly determined domain configurations at different stages of the polarization reversal process. Recently, we further improved this PFM approach by adopting a separate probe needle that can be used to apply an external bias field and *ac* modulation field through a top electrode and to obtain a polarization switching current. In this setup the PFM tip was used to detect only piezoresponse signals. Using this modified PFM method, we could visualize the inhomogeneous domain nucleation in epitaxial FE capacitors.[12]

Here, we report that the modified PFM method can provide a reliable way to obtain *t*-dependent domain evolution images. By comparing the switched polarization ($\Delta P$) values from the PFM images with those from the switching current measurements, we can avoid possible errors due to imprint, relaxation, and other reliability issues. From the PFM images, we determined the domain wall speed (*v*) and observed decrease in *v* with increases in domain size. In addition, we observed the average value of *v* showed creep behavior, *i.e.* <*v*> ~ $\exp(-E_0/E_{app})^\mu$.[4,5]

A highly-polar epitaxial $PbZr_{0.2}Ti_{0.8}O_3$ (PZT) film ($P_r \approx 80$ $\mu C/cm^2$) grown on a conducting $SrRuO_3$ electroded $SrTiO_3$ (001) substrate by pulsed laser deposition was used for this study.[16] High resolution *x*-ray diffraction confirmed that the 100 nm-thick PZT epitaxial film is purely *c*-axis oriented without contribution from *a*-axis oriented domains or impurity phases. Pt top electrodes (100 $\mu$m in diameter) were deposited. For the modified PFM measurements, we used the experimental setup shown in Fig. 1 of Ref. 12. We measured amplitude (*R*) and phase ($\theta$) of the piezoelectric signals, and the volume fraction (*q*) of the reversed domain was obtained by integrating the $R\cos\theta$ values.[12]

We found that the experimental procedures, *i.e.* pulse sequences and PFM image scans, are very important to obtain reliable information on domain wall motions. Figure 1(a) shows the experimental procedure for the step-by-step switching approach suggested by Gruverman *et al.*[10] It is highly desirable to obtain successive PFM images of domain wall evolution that occurs during a single polarization switching process. However, the experimental procedure shown in Fig. 1(a) cannot provide such successive PFM images as each poling pulse resets the capacitor into the fully polarized state. Even when we use switching pulses with the same pulse width (*i.e.* $t_1 = t_2$), the obtained PFM images will be somewhat different due to the stochastic nature of inhomogeneous nucleation.[12]

To circumvent this difficulty, we adopted a different experimental procedure, shown in Fig. 1(b). After applying one poling pulse (10 V, 50 $\mu$s), we applied a series of pulses with pulse width given by $\tau_1, \tau_2, \tau_3, \ldots, \tau_n$, and measured the PFM images between the pulses. We assumed that *the PFM image obtained after the i-th pulse could be nearly the same as that after a single pulse of $t = \tau_1 + \tau_2 + \ldots + \tau_i$*. Then all the PFM images taken before the (*i*+1)-th pulse could reveal the successive domain wall evolution due to the polarization switching during the single pulse of *t*. Note that our assumption will be valid only if domain walls do not move during the time gap between the applied pulses, especially during PFM scanning which typically takes several minutes. Many ferroelectric films have reliability problems, such as imprint and relaxation.[1] When such problems occur, the PFM images obtained from Fig. 1(b) cannot correctly represent the successive domain wall motion during the single pulse.

To confirm the reliability of the obtained images, we performed switching current measurements independently, just after taking the successive PFM images. For switching current measurements, we used the pulse sequence shown in Fig. 1(a) but without PFM scanning, and obtained $\Delta P$.[3] Note that the normalized switched polarization ($\Delta p \equiv \Delta P/2P_r$) should be the same as $q$ from the PFM images if our assumption holds true, *i.e.* domain walls will not move during the PFM studies in Fig. 1(b).

First, we measured $\Delta p$ and $q$ values for a 45 nm-thick epitaxial PZT film, which showed polarization relaxation caused by the depolarization field, *i.e.* a decrease of polarization over time without $E_{app}$.[17,18] The solid squares in Fig. 1(c) show the difference between $\Delta p$ and $q$. Only after $t = 5$ $\mu$s, ($\Delta p - q$) becomes larger than 16 %. This disagreement indicated that our modified PFM studies with the pulse sequences in Fig. 1(b) cannot provide correct information on domain evolution in the 45 nm-thick PZT film.

On the other hand, we repeated the same experiments using our 100 nm-thick epitaxial PZT film, which did not show any relaxation or retention problems from other electrical measurements. The open circles in Fig. 1(c) denote ($\Delta p - q$) values, which remain less than 3 % up to $t = 10$ $\mu$s. This good agreement between $\Delta p$ and $q$ demonstrated the validity of our assumption for our 100 nm-thick PZT film. In the remainder of this paper, we will focus on our experimental data for the 100nm-thick PZT film only.

With this reliable experimental procedure using modified PFM, we could investigate domain wall motion from successive PFM images. Figure 2 shows $R\cos\theta$ images obtained at the same sample position under different voltages ($V_{app}$). By comparing the $R\cos\theta$ images, we were able to observe all the steps of the domain switching process, *i.e.* nucleation, forward growth, sidewise domain wall motion, and coalescence of reversed domains. The arrow in the image of $V_{app}$ = -3 V and $t$ = 30 µs shows a site with a newly formed nucleus with reversed polarization. The image contrast of nuclei increased with time, indicating forward growth of the reversed polarization. After its contrast reached a maximum, the reversed domain began to expand sidewise and finally merged with another reversed domain, as shown in Fig. 3(a). By comparing the images of $V_{app}$ = -3 V, -6 V, and -8 V in Fig. 2, we can easily see that nucleation is much more important in the early stage of switching and under a high voltage.[12]

We could determine *t*-dependent domain wall speed *v* quantitatively from the successive PFM images. Figure 3(a) shows the *t*-dependent evolution of reversed domains under a bias of -6 V, for the region marked by a dashed box in Fig. 2. To estimate *v* values, we approximated the domain near the center position as having a circular shape with a mean radius (*r*). The value of *v* between $t_1$ and $t_2$ was obtained from $v\,((t_2+ t_1)/2) = (r(t_2) – r(t_1)) / (t_2 – t_1)$. We stopped estimation of the *v* values when a domain merged with another domain, as shown in the image at $t$ = 8 µs in Fig. 3(a). Figure 3(b) shows the domain area and *v* of the reversed domain as a function of *t*. As *t* increased, *v* decreased.

To obtain further insights, we measured *v* values for many reversed domains and found an interesting dependence of *v* on domain size. Figure 4(a) shows the log-log plot of *v* and *r*, obtained from the PFM images under a bias of -8 V. It was evident that *v* decreased as *r* increased. According to the classical theory of phase-ordering kinetics, curvature of the phase boundary plays a key role in modifying *v*. Surface tension, which is proportional to the mean curvature, will act at each point on the domain wall and cause the domain wall to move with $v \sim r^{-1}$ even without $E_{app}$.[19] However, in our PZT film, the domain walls did not move without $E_{app}$, indicating that they were pinned by internal disorder. Therefore, the observed domain wall growth behavior should be related more closely to motion of interfaces in a disordered medium.

Motion of interfaces in a disordered medium at finite temperature can be described as a creep motion in a low $E_{app}$ regime. Then $<v>$ should be proportional to $\exp(-E_0/E_{app})^{\mu}$, where $E_0$ and $\mu$ are an activation field, and a dynamical exponent, respectively. As $v$ depends on the domain size, we obtained the $<v>$ values by averaging $v$ for numerous domains of nearly circular shape with $q$ values between 0.2 and 0.4. Figure 4(b) shows the log-log plot of $<v>$ and $1/E_{app}$. $<v>$ increased as $E_{app}$ increased. The solid (red) line shows the fitting curve, which gave the values of $\mu \sim 0.9 \pm 0.1$ and $E_0 \sim 700$ kV/cm. Note that our data is also consistent with recent report on creep behavior,[4] which correspond to $\mu = 1.0$.

Recently, the FE domain wall growth behaviors have attracted lots of attention due to their scientific and technological importance. For examples, the FE domain wall was reported to have a fractal dimension.[5,13] And importance of the collective effect of defects on domain switching was also addressed.[8,9] Such intriguing phenomena should be also closely related to the motion of domain walls in the disordered medium, so further systematic studies are highly desirable.

In summary, we investigated the time-dependent domain wall motion of epitaxial Pb(Zr,Ti)O$_3$ thin films by modified piezoresponse force microscopy combined with switching current measurements. We found that our developed PFM method can be applied to obtain successive microscopic images of domain evolution during the polarization switching process. Our experimental technique will provide a useful tool to investigate domain wall dynamics in a capacitor geometry, which is commonly used in most FE devices.

This study was financially supported by the Creative Research Initiative program

(Functionally Integrated Oxide Heterostructure) of MOST/KOSEF. S. M. Y. acknowledges financial support, in part, from the Seoul Science Scholarship. H. N. L. was supported by the LDRD program of ORNL.

**Figure captions**

Fig. 1. (Color online) Schematic diagrams of pulse sequences used (a) to employ the step-by-step switching approach and (b) to obtain successive PFM images for our PFM measurements. To estimate the reliability of our experiments, we evaluated the volume fraction ($q$) of the reversed domain obtained by PFM measurements and compared the results with normalized switched polarization ($\Delta p$) from the switching current measurements *i.e.* using the pulse sequence in (a) but without the PFM scanning. (c) Values of ($\Delta p - q$) for 45 nm and 100 nm PZT films, which did and did not have relaxation problems, respectively.

Fig. 2. (Color online) Visualization of successive domain evolution images ($R\cos\theta$) at various $t$ at $V_{app}$ = -3 V, -6 V, and -8 V. The scan area was $5 \times 5$ $\mu m^2$. Contrast indicates the amount of reversed polarization, *i.e.* $R\cos\theta$.

Fig. 3. (Color online) (a) Successive $t$-dependent evolution images of a reversed domain under -6 V bias for the region marked by a dashed box in Fig. 2. (b) Plots of domain wall speed ($v$) and domain area as a function of $t$.

Fig. 4. (Color online) (a) Log-log plot of domain wall speed ($v$) and domain radius ($r$) under -8 V bias. The dashed (blue) line corresponds to the theoretical prediction, *i.e.* $v \sim r^{-1}$, based on surface tension. And the solid (red) line is only for guideline of eye. (b) Averaged domain wall speed ($<v>$) as a function of the inverse applied electric field ($1/E_{app}$). The linear (red) line indicates the fitting to the creep motion, *i.e.* $<v> \sim \exp(-E_0/E_{app})^\mu$.

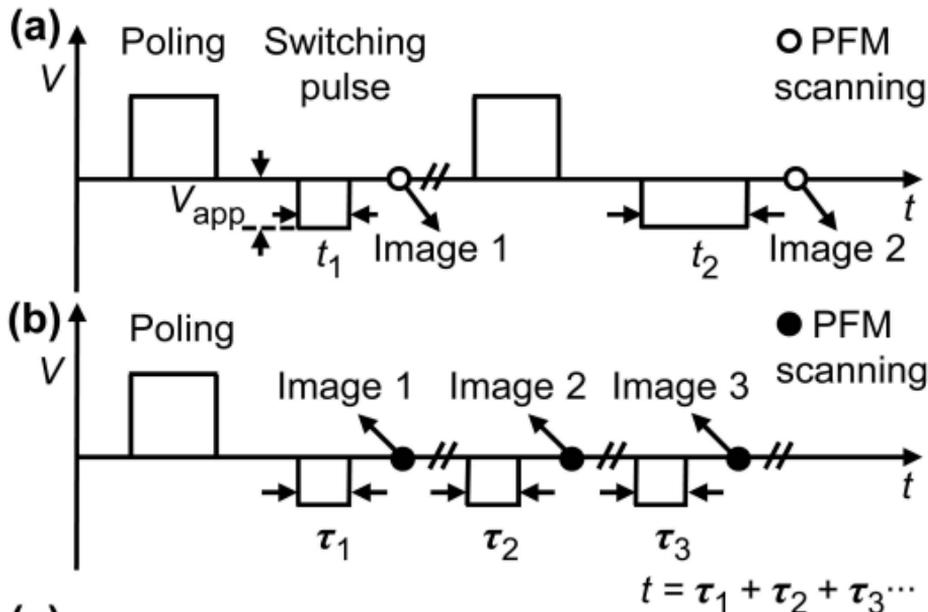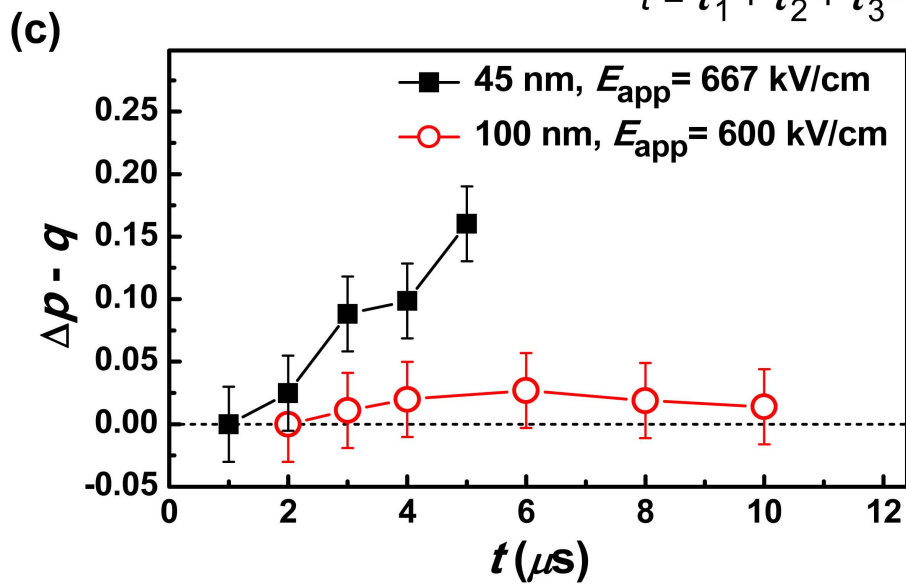

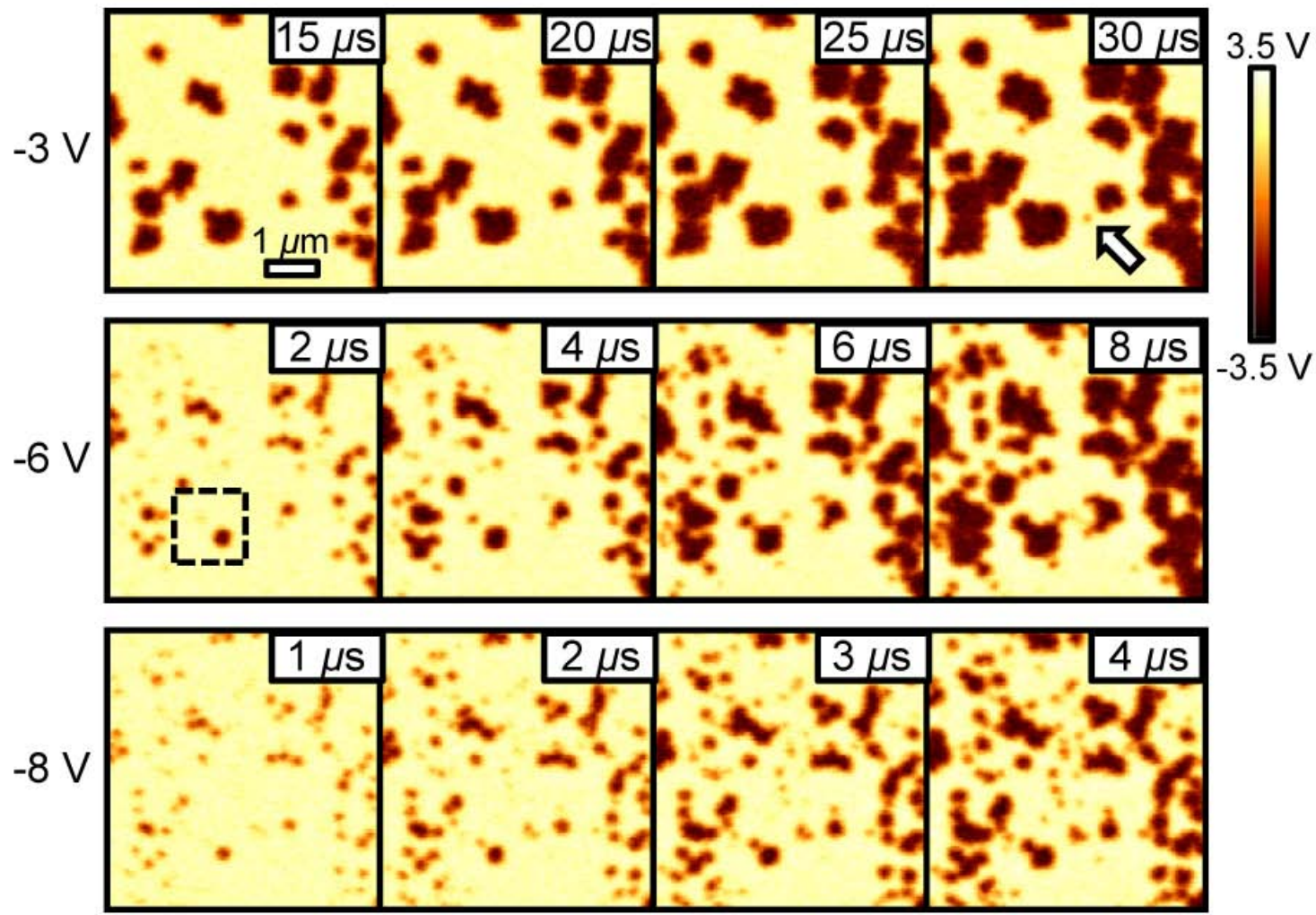

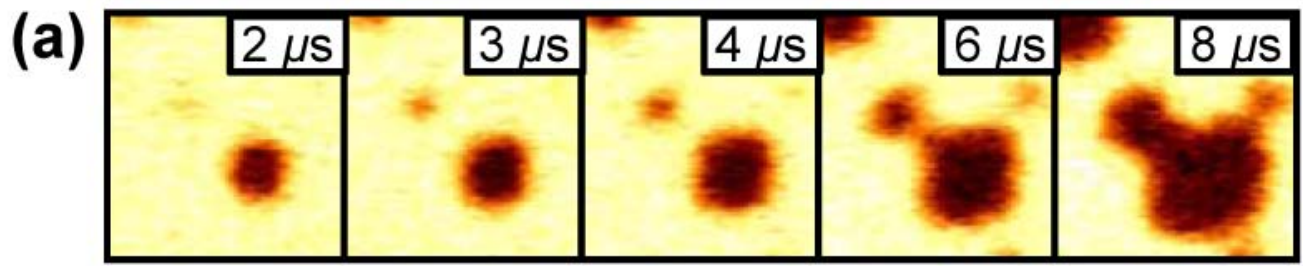
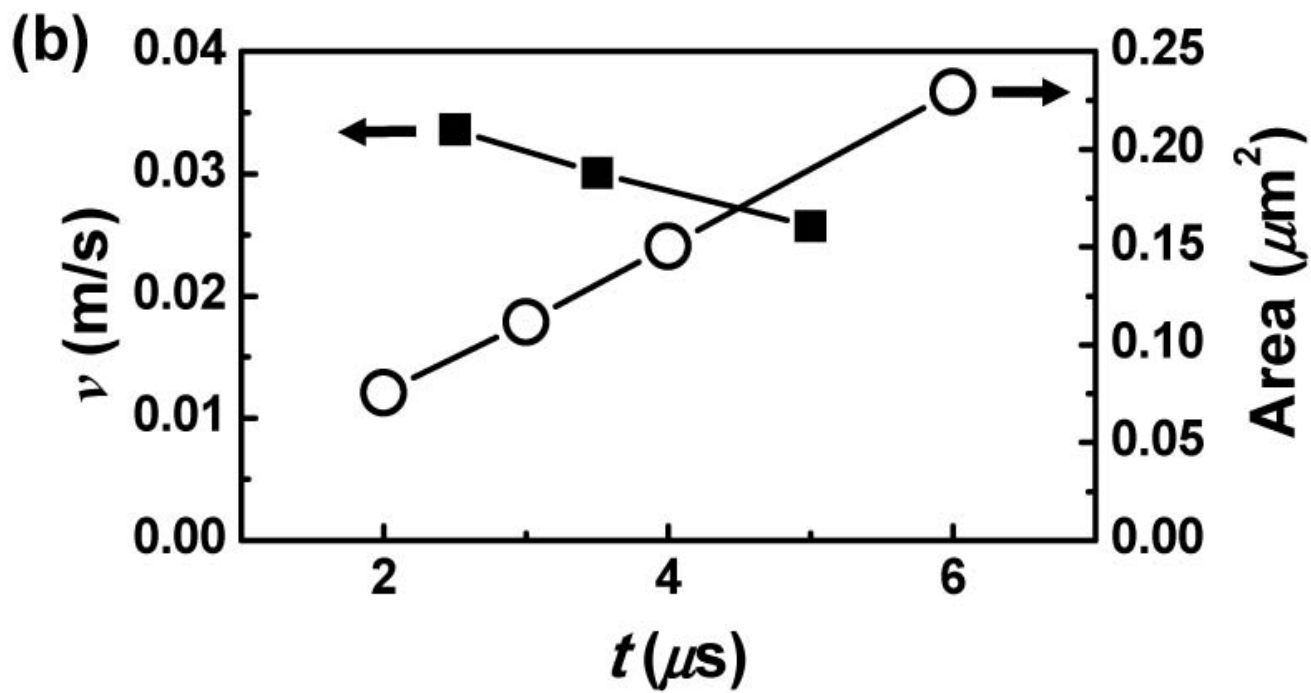

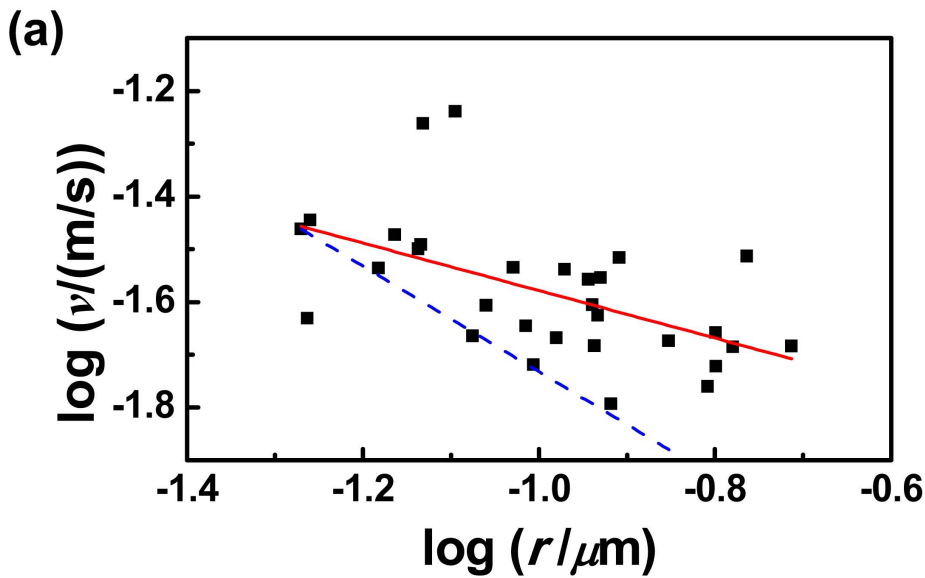

(a)

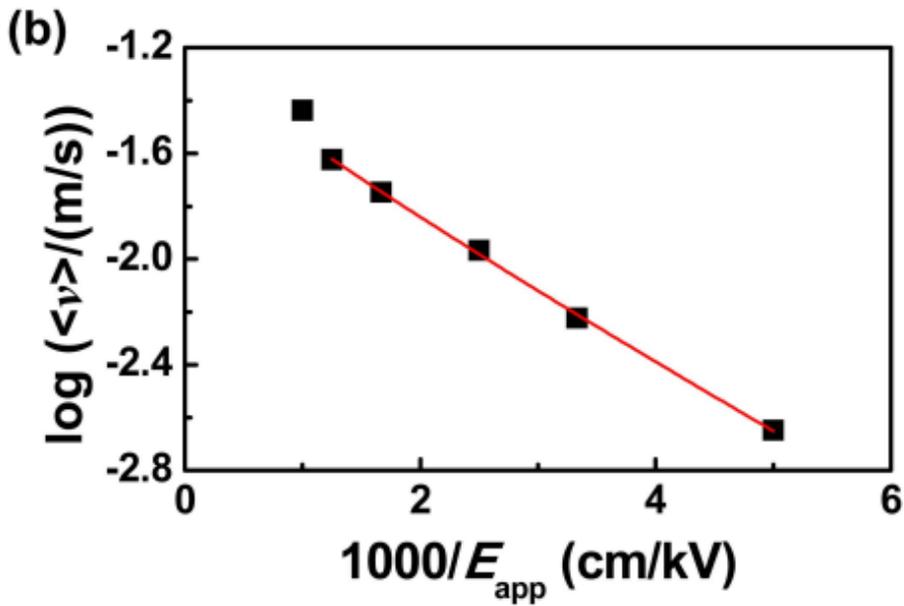

(b)